\documentclass[showpacs,twocolumn,showkeys,amsmath,amssymb,prl,nofootinbib
]{revtex4-1}

\usepackage{amsmath,amssymb,amsfonts}
\usepackage{graphicx}
\usepackage{txfonts}
\usepackage{epstopdf}
\usepackage[T1]{fontenc}

\usepackage{color}

\usepackage[unicode]{hyperref}
\hypersetup{
   unicode=true,          
   a4paper=true,
   plainpages=false,
   pdftitle={Title of PDF},    
   pdfauthor={Author of PDF},     
   pdfsubject={Subject of PDF},   
   colorlinks=true,       
   citecolor=blue,        
}
\begin{document}

\title{Finite-temperature phase transition in a homogeneous one-dimensional gas of attractive bosons
}

\author{Christoph Weiss}
\email{christoph.weiss@durham.ac.uk}
  \affiliation{Joint Quantum Centre (JQC) Durham--Newcastle, Department of Physics, Durham University, Durham DH1 3LE, United Kingdom}

\date{Submitted: August 16, 2013; current version October 29, 2016}

 \begin{abstract} 
In typical one-dimensional models the Mermin-Wagner theorem forbids long range order, thus preventing finite-temperature phase transitions. We find a finite-temperature phase transition for a homogeneous system of attractive bosons in one dimension. The low-temperature phase is characterized by a quantum bright soliton without long range order; the high-temperature phase is a free gas.  Numerical calculations for finite particle numbers show a specific heat scaling as $N^2$, consistent with a vanishing transition region in the thermodynamic limit.
\end{abstract}
\pacs{05.70.Fh,
03.75.Lm,
05.30.Jp,
03.75.Hh 
}

\keywords{phase transition, one dimension, finite temperatures, bright solitons, Bethe ansatz, Lieb-Liniger model, attractive interactions, Mermin-Wagner theorem, Bose-Einstein condensation}

\maketitle

Bright solitons generated from attractively interacting Bose-Einstein condensates in quasi-one-dimensional wave guides are investigated experimentally in an increasing number of experiments\cite{KhaykovichEtAl2002,StreckerEtAl2002,CornishEtAl2006,MarchantEtAl2013,MedleyEtAl2014,McDonaldEtAl2014,NguyenEtAl2014,MarchantEtAl2016,EverittEtAl2015,LepoutreEtAl2016}. As experiments do not truly take place in one dimension but rather in \textit{quasi}-one-dimensional wave guides, providing a thermalization mechanism~\cite{MazetsSchmiedmayer2010,CockburnEtAl2011}, this leads to the question whether or not these bright solitons can be stable in the presence of thermal fluctuations.

The Mermin-Wagner theorem~\cite{MerminWagner1966} proves that in many models long-range order in one or two dimensions cannot exist at finite temperatures~\cite{MerminWagner1966,Hohenberg1967}; this excludes the existence of many phase transitions. {Finite-temperature transitions are fundamentally different from quantum phase transitions (cf.~\cite{JakschEtAl1998,GreinerEtAl2002}); one-dimensional quantum phase transitions can be found, e.g., in Refs.~\cite{Zwerger2003,CincioEtAl2007,KanamotoEtAl2010}).} While there are some finite temperature phase transitions in low-dimensional systems  like the Berezinsky-Kosterlitz-Thouless transition in two dimensions~\cite{HadzibabicEtAl2006}
 or the phase transition in the two-dimensional Ising model~\cite{Onsager1944}, the generic case is that low-dimensional models to not undergo finite-temperature phase transitions~\cite{GelfertNolting2001}. Indeed, a book on ``thermodynamics of one-dimensional \textit{solvable}\/ models'' does not include the word {``phase transition''}\/ in its index~\cite{Takahashi2005}. For a \textit{disordered}\/ system displaying Anderson-localization~\cite{Anderson1958}, a finite-temperature phase transition for weakly interacting bosons in one dimension has been found in Ref.~\cite{AleinerEtAl2010}. 

A quasi one-dimensional system of attractively interacting bosons can be modeled~\cite{LaiHaus1989,CastinHerzog2001,CalabreseCaux2007,MuthFleischhauer2010} by the \textit{solvable}\/ Lieb-Liniger model~\cite{LiebLiniger1963,McGuire1964,SeiringerYin2008}. 
One of the challenges for bright-soliton experiments~\cite{KhaykovichEtAl2002,StreckerEtAl2002,CornishEtAl2006,Hulet2010b,MarchantEtAl2013,MedleyEtAl2014,McDonaldEtAl2014,NguyenEtAl2014} is to realize true quantum behavior predicted, so far, with zero-temperature calculations~\cite{CarrBrand2004,WeissCastin2009,StreltsovEtAl2009b,SachaEtAl2009,StreltsovEtAl2011,GertjerenkenEtAl2013,BarbieroSalasnich2014}. For the Lieb-Liniger model, investigations of thermal effects on the many-body level for bosons in one dimension have so far focused on the more extensively studied case of repulsive interactions (Ref.~\cite{Takahashi2005} and references therein); for finite systems classical field methods have been applied~\cite{BieniasEtAl2011}. In other soliton models, thermodynamics with interacting solitons has been investigated~\cite{ZabuskyKruskal1965,CurrieEtAl1980}.

In this Letter we show that attractive bosons in the Lieb-Liniger model undergo a finite-temperature phase transition; a bright soliton -- no-soliton transition. As bright solitons do not display long-range order, this does not violate the Mermin-Wagner theorem.
Although bright solitons do not display long-range order, quantum bright solitons are fundamentally different from localized states cf.~\cite{AleinerEtAl2010}:
For the Lieb-Liniger model, the energy eigenfunction describing a soliton of $N$-particles has to obey the symmetry of the Hamiltonian and is thus translationally invariant.

For $N$ identical bosons on a one-dimensional line of length $L$, corresponding to the experimentally realizable~\cite{SchmidutzEtAl2014} box potential, the Lieb-Liniger Hamiltonian reads~\cite{LiebLiniger1963,McGuire1964,SeiringerYin2008}
\begin{align}
\nonumber
\hat{H} = &-\sum_{j=1}^{N}\frac{\hbar^2}{2m}\frac{\partial^2}{\partial{x_j}^2} + \sum_{j=1}^{N-1}\sum_{n=j+1}^{N}g_{1\rm D}\delta\left(x_j-x_{n}\right), 
\end{align}
where $g_{1\rm D} < 0$ quantifies the contact interactions between two particles,   $m$ is the mass,  and $x_j$ the position of the $j$th particle. 
Contrary to the phenomenological model used in~\cite{DunjkoEtAl2003}
 for a harmonically trapped one-dimensional gas of attractive bosons, we use the complete set of energy-eigenvalues which are known analytically for large $L$\footnote{The precise limit, which was not discussed in Refs.~\cite{LiebLiniger1963,McGuire1964}, will be defined in Eq.~(\ref{eq:limit}) after the necessary physical requirements on this limit are stated.}~\cite{CastinHerzog2001,SykesEtAl2007},
\begin{equation}
\label{eq:eigenenergies}
E_{\rm LL}\left(\big\{n_{r},k_r\big\}_{r=1\ldots R}\right) =
\sum_{r=1}^R\left(E_0(n_r)+\frac{\hbar^2K_r^2}{2 n_r m}\right),\quad \sum_{r=1}^Rn_r=N,
\end{equation}
where the $R$ natural numbers  $n_r$ correspond to either free particles, if $n_r=1$, or matter-wave bright solitons, if $n_r>1$ (cf.\ the energy-eigenfunctions discussed in Ref.~\cite{CastinHerzog2001}; for experiments with more than two solitons see Refs.~\cite{StreckerEtAl2002,CornishEtAl2006}, cf.~\cite{CarrBrand2004,StreltsovEtAl2011}). Each soliton has kinetic energy (proportional to the square of the single-particle momentum $\hbar k_r$, shared by all particles belonging to this soliton) and ground-state energy~\cite{McGuire1964,CastinHerzog2001}
\begin{equation}
\label{eq:E0}
E_0(n_r) =-\frac1{24}\frac{mg_{1\rm D}^2}{\hbar^2}n_r(n_r^2-1).
\end{equation}
We choose periodic boundary conditions (cf.~\cite{SykesEtAl2007}) which lead to $KL$ having to be an integer multiple of $2\pi$, thus
\begin{equation}
\nonumber
K_r = \frac{2\pi}L \nu_r,\quad \nu_r = \ldots,-2,-1,0,1,2,\ldots
\end{equation}
As we are dealing with indistinguishable particles, many-particle wave functions~\cite{CastinHerzog2001} are unambiguously defined by only considering configurations with
\begin{equation}
\nonumber
n_1\ge n_2\ge n_3 \ge \ldots \ge n_R.
\end{equation}
Because of Eq.~(\ref{eq:eigenenergies}), the total number of possibilities to distribute $N$ particles among up to $N$ parts is thus given by the number partitioning problem~\cite{AbramowitzStegun1984}
\begin{equation}
p(N)\sim \frac 1{4 N \sqrt{3}}\exp\left(\frac{\pi\sqrt{2}}{\sqrt{3}}\sqrt{N}\right),\quad N\gg 1\;.
\label{eq:exgrowth}
\end{equation}

The $\propto N^3$-dependence of the ground-state energy~(\ref{eq:E0}) is a problem for the treatment of the thermodynamic limit ($N\to\infty$, $L\to\infty$ such that $N/L=\rm const.$)~\cite{Takahashi2005}; the Lieb-Liniger model with repulsive interaction is thus normally used to do thermodynamics~\cite{Takahashi2005}. However, for attractive interactions treating the limit $N\to \infty$ at fixed interaction would lead to infinite densities, cf.~\cite{CastinHerzog2001}. We thus combine the thermodynamic limit with vanishing interaction -- as used in the mean-field (Gross-Pitaevskii) theory of bright solitons~\cite{PitaevskiiStringari2003}.
\begin{align}
\label{eq:limit}
N\to\infty,\quad L\to\infty,\quad g_{\rm 1D}\to 0, 
\quad \varrho = \textrm{const.},\quad \widetilde{g} = \textrm{const.},
\end{align}
where 
$\varrho \equiv N/L$ and $\widetilde{g} \equiv Ng_{\rm 1D}$.
When approaching the limit~(\ref{eq:limit}), the energy-gap $E_{{\rm gap}}\equiv E_0(N-1)-E_0(N)$ is an $N$-independent energy scale which will turn out to be the relevant energy scale for thermodynamics; we can express characteristic temperatures as
\begin{equation}
\label{eq:Tdef}
 k_{\rm B} T_0 = AE_{{\rm gap}};
\end{equation}
and subsequently investigate if the prefactor $A$ remains non-zero in the limit~(\ref{eq:limit}).
The ground state energy~(\ref{eq:E0}) now reads
\begin{align}
\nonumber
E_0(n_r) &= -\frac{E_{{\rm gap}}}{3N(N-1)}n_r(n_r^2-1);
\end{align}
in the limit (\ref{eq:limit}) the energy gap is given by
\begin{align}
\nonumber
 E_{{\rm gap}} &= \frac18\frac{m\widetilde{g}^2}{\hbar^2}={\rm const.} >0.
\end{align}
 Before we choose the \textit{canonical ensemble}\/ (characterized by temperature~$T$ and particle number~$N$~\cite{PathriaBeale2011}) to do thermodynamics, we should quantify the requirement that $L$ has to be large in order for the energy-eigenvalues~(\ref{eq:eigenenergies}) to be correct within the limit~(\ref{eq:limit}). The ground-state wave function for $N$ bosons is given by 
$\psi_0\propto \exp[- m|g_{1\rm D}|/(2\hbar^2)\sum_{1\le j<n\le N}|x_j-x_n|]$; the size of an $N$-particle soliton $\sigma\propto 1/(|g_{\rm 1D}|N)$~\cite{CastinHerzog2001} and thus remains a non-zero constant in the limit~(\ref{eq:limit}), leading to a single particle density $\propto \cosh(x/\sigma)^{-2}$ and thus also to a vanishing off-diagonal long-range order.\footnote{\label{footnote:order}The many-particle ground state can be viewed as consisting of a relative wave-function given by a Hartree product state with $N$ particles occupying the GPE-soliton mode $\propto \cosh[(x-x_0)/\sigma]^{-1}$ and a center-of-mass wave function for the variable $x_0$ (cf.~\cite{CalogeroDegasperis1975,CastinHerzog2001}). The one-body density matrix~\cite{PitaevskiiStringari2003} then is $\propto \cosh[(x-x_0)/\sigma]^{-1}\cosh[(x'-x_0)/\sigma]^{-1}$ which vanishes in the limit $|x-x'|\to\infty$ even after integrating over $x_0$. Thus, there is no off-diagonal long range order in our system.}
In order for the energy eigenvalues given by Eq.~(\ref{eq:eigenenergies}) to be valid, the system has to be larger than the size of a $N=2$ soliton (the more particles are in a soliton, the smaller it gets~\cite{CastinHerzog2001}). To be on the safe side we ask the wave function to be below $e^{-100}$ for particle separation greater than $L$, that is 
\begin{equation}
\nonumber
\frac{m|g_{1\rm D}|}{2\hbar^2}L\gtrapprox 100.
\end{equation}
For the two relevant energy scales of Eq.~(\ref{eq:eigenenergies}) this gives an energy ratio
\begin{align}
\nonumber
\mathcal{E}(N) &\equiv \frac{E_{\rm gap}}{E_{n_r=1,\rm kin}(\nu_r=1)} 
= B N^2,\\  B&\equiv\left(\frac{mg_{\rm 1D}}{2\hbar^2}L\right)^2\frac{1}{(2\pi)^2};
\label{eq:ratio}
\end{align}
the eigenvalues~(\ref{eq:eigenenergies}) are therefore a very good approximation to the true eigenvalues of the Lieb-Linger model (for all temperatures) if
\begin{equation}
B \gtrapprox B_0\equiv \frac{100^2}{(2\pi)^2}\simeq 253.
\end{equation}

For any choice of $\{n_r\}_{r=1..R}$, the canonical partition function will depend on how often solitons of exactly size $n_r$ occur. We thus rewrite these configurations, now listing them using distinct integers $n'_{r}$ with $n'_{r}>n'_{r+1}$ and the multiplicity $\#(n'_{r})$ with which the value  $n_r$ had occurred:
\begin{equation}
\nonumber
\{n_r\}_{r=1..R} \longrightarrow \left\{\big(n_r',\#(n_r')\big)\right\}_{r=1..R'},\quad\sum_{r=1}^{R'}n_r'\#(n_r')=N.
\end{equation}
Note that replacing $\{n_r\}_{r=1..R}$ by $\left\{\big(n_r',\#(n_r')\big)\right\}_{r=1..R'}$ is bijective, that is, to each set of $n_r$ there is exactly one set of 
$\big(n_r',\#(n_r')\big)$ (and vice versa); in the following we can thus always use the notation which is more convenient.
The total canonical partition function is the sum 
\begin{equation}
\label{eq:ZNtotal}
Z_{N,\rm total}(\beta)\equiv \sum_{\substack{\{n_{r}\}_{r=1..R} \\\sum_{r=1}^{R}n_r=N}}Z_{N,\left\{\big(n_r',\#(n_r')\big)\right\}_{r=1..R'}}(\beta)
\end{equation}
over the partition functions for fixed $\{n_r\}_{r=1..R}$ 
\begin{equation}
\label{eq:ZNchosenconfig}
Z_{N,\left\{\big(n_r',\#(n_r')\big)\right\}_{r=1..R'}}(\beta)=\prod_{r=1}^{R'}e^{-\#(n_r')\beta E_0(n_r')} Z_{n_r',\#(n_r'), \rm kin}(\beta),
\end{equation}
where the kinetic part can be calculated using the recurrence relation~\cite{Landsberg1961} (which has been used to describe ideal Bose gases, e.g., in Refs.~\cite{BrosensEtAl1996,WeissWilkens1997,GrossmannHolthaus1997b})
\begin{align}
\label{eq:Zrec}
Z_{n_r,\#(n_r), \rm kin}(\beta)&=\frac1{\#(n_r)}\sum_{\ell=1}^{\#(n_r)}Z_{n_r,1, \rm kin}(\ell \beta)Z_{n_r,\#(n_r)-\ell, \rm kin}(\beta),
\end{align}
with $ Z_{n_r,0, \rm kin}(\beta)\equiv 1$
and the kinetic energy part of the single-soliton partition function is given by\footnote{When approximating the sum $\sum_{\nu=-\infty}^{\infty}\exp(-x\nu^2)$ by the integral  $\int_{-\infty}^{\infty}d\nu\;\exp(-x\nu^2)$, the error lies below $10^{-40}$ for $0<x<0.1$~\cite{maple}. When approaching the limit~(\ref{eq:limit}), $x\to 0$ and Eq.~(\ref{eq:Zonesoliton}) thus becomes exact.} 
\begin{align}
Z_{n_r,1, \rm kin}(\beta)
&=\sum_{\nu=-\infty}^{\infty}\exp\left(-\beta \frac{E_{\rm gap}}{n_r BN^2}\nu^2\right)\nonumber\\
&\simeq \int_{-\infty}^{\infty}d\nu\;\exp\left(-\beta \frac{E_{\rm
      gap}}{n_rBN^2}\nu^2\right)=\left(\frac{\pi n_r BN^2}{\beta E_{\rm gap}}\right)^{\frac 12}.
\label{eq:Zonesoliton}
\end{align}
Rather than having to explicitly do sums over a large number~(\ref{eq:exgrowth}) of configurations, for larger particle numbers it is preferable to calculate the partition function again via a recurrence relation, starting with $R=1$ and $Z^{(R=1)}_{M,n_R,\#(n_R)}(\beta)$, $M=1,2,\ldots N$ given by Eq.~(\ref{eq:ZNchosenconfig}). The step $R\to R+1$ then yields the case $n_{R+1}=n_{R}$
 with
\begin{align}
 Z^{(R+1)}_{M+n_{R+1},n_{R+1},\#(n_{R+1})+1}(\beta)=&\frac{e^{-\beta E(n_{R+1})} Z_{n_{R+1},\#(n_{R+1})+1, \rm kin}(\beta)}{Z_{n_{R+1},\#(n_{R+1}), \rm kin}(\beta)}\nonumber \\
&\times Z^{(R)}_{M,n_{R+1},\#(n_{R+1})}(\beta)
\label{eq:exactrec1}
\end{align}
as well as
\begin{align}
 Z^{(R+1)}_{M+n_{R+1},n_{R+1},1}(\beta)
 =
 ~&e^{-\beta E(n_{R+1})}Z_{n_{R+1},1, \rm kin}(\beta)
\nonumber \\
&\times\sum_{n_R=n_{R+1}+1}^{M}\sum^{\lfloor M/n_R\rfloor}_{{\#(n_R)=1}}Z^{(R)}_{M,n_R,\#(n_R)}(\beta),
\label{eq:exactrec2}
\end{align}
where $\lfloor x\rfloor$ denotes the largest integer $\le x$.

From the total canonical partition function~(\ref{eq:ZNtotal}) we obtain the specific heat (at fixed particle number $N$ and system size $L$, which is proportional to the variance of the energy) as
\begin{align}
\label{eq:C}
C_{N,L}(T) &\equiv \frac{\partial}{\partial T}\langle E\rangle=-\frac{\partial}{\partial T}\frac{\partial}{\partial \beta}\ln[Z_{N, \rm total}(\beta)]\\ &= \frac1{k_{\rm B}T^2}\frac{\partial^2}{\partial \beta^2}\ln[Z_{N, \rm total}(\beta)]  =  \frac1{k_{\rm B}T^2}\left(\langle E^2\rangle-\langle E\rangle^2\right);
\nonumber
\end{align}
the number of atoms in the largest soliton is given by
\begin{equation}
\label{eq:n1}
\langle n_1\rangle(T) = \frac1{Z_{N, \rm total}(\beta)}\sum_{\substack{\{n_{r}\},r=1..R \\\sum_{r=1}^{R}n_r=N}}n_1Z_{N,\left\{\big(n_r',\#(n_r')\big)\right\}_{r=1..R'}}(\beta).
\end{equation}
For analytic calculations Eq.~(\ref{eq:Zonesoliton}) leads to 
\begin{align}
\frac1{[\#(n_r)]!}\left(\frac{\pi n_r BN^2}{\beta E_{\rm gap}}\right)^{\frac{\#(n_r)}2}
 \le
Z_{n_r,\#(n_r), \rm kin}(\beta)\nonumber\\
\le \frac{e^{\#(n_r) c_1}}{[\#(n_r)]!}\left(\frac{\pi n_r BN^2}{\beta E_{\rm gap}}\right)^{\frac{\#(n_r)}2},\quad c_1\equiv\ln(2),
\label{eq:bounds}
\end{align}
the lower bound being (for temperatures large compared to the center-of-mass first exited state) the largest term involved in the sum~(\ref{eq:Zrec}); to obtain the upper bound we choose the value for $c_1$ such that all $2^{\#(n_r)-1}<e^{\#(n_r)\ln(2)}$ addends in the sum~(\ref{eq:Zrec}) (treated separately) are of the same order as the highest term.

\begin{figure}
\includegraphics[width=\linewidth]{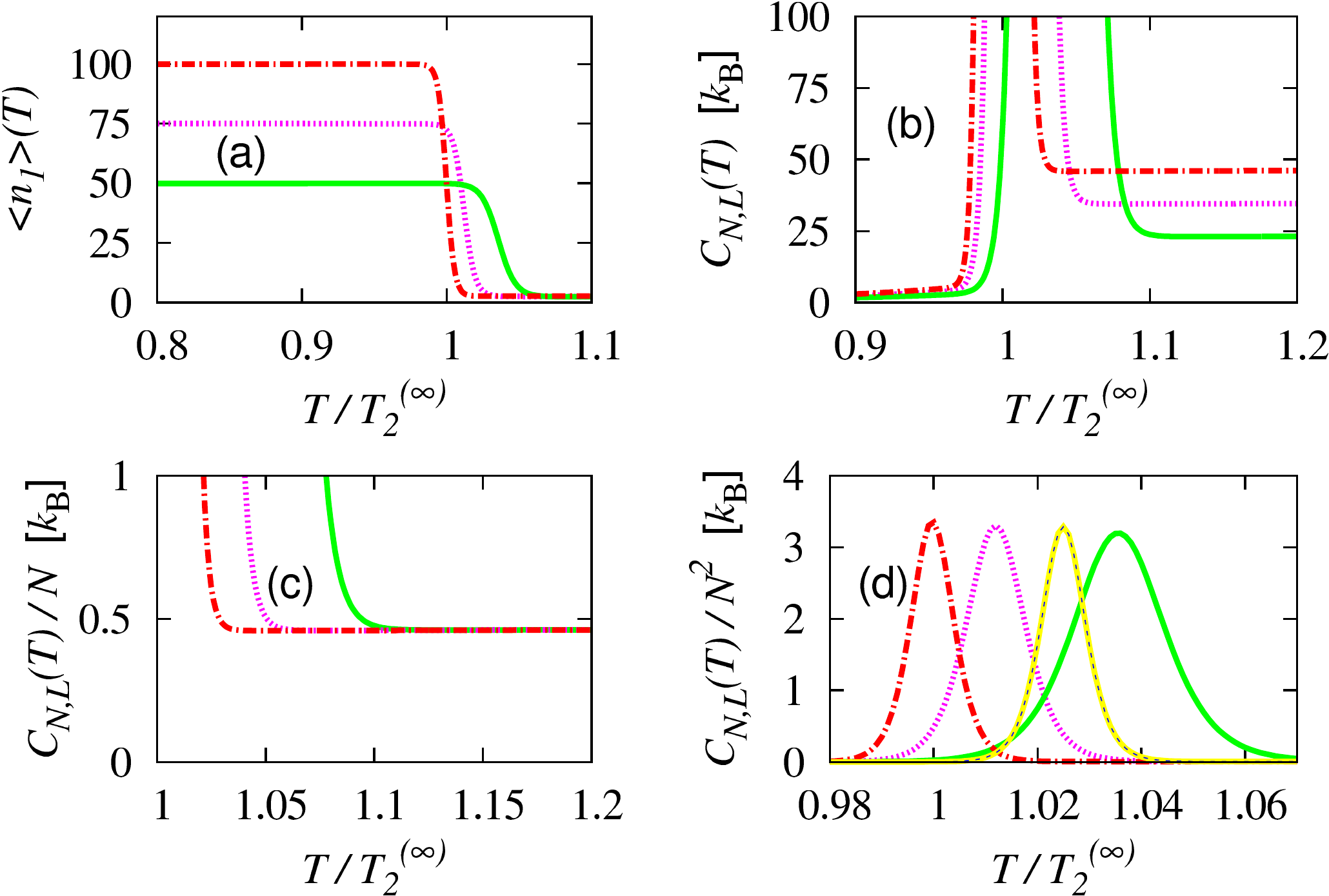}
\caption{\label{fig:transition}(Color online)
Finite size investigations of the soliton--non-soliton transition for $N= 50$ (green/gray solid line), $N=100$ (red/black dash-dotted line) and $N=75$ (magenta/gray dashed line) with $B=B_0$. (a) Size of the largest soliton~(\ref{eq:n1}) as a function of temperature.   (b) For $N\approx 100$, the specific heat~(\ref{eq:C}) as a function of temperature scales as $0.5 k_{\rm B}$ at low temperatures indicating one bright soliton. (c) At high temperatures, it scales as  $\lessapprox 0.5N k_{\rm B}$, demonstrating a free gas of $N$ atoms. (d) The specific heat near the transition temperature scales as $N^2k_{\rm B}$. Excluding all states with more than one soliton (yellow/light gray line, $N=100$) clearly indicates the presence of several solitons near the transition temperature; the area of coexistence gets smaller for increasing $N$.}
\end{figure}

In order to define a characteristic temperature~(\ref{eq:Tdef}), we now use the temperature below which finding a single soliton with $N$ particles is more probable than finding $N$ single particles. Both partition functions, evaluated at $T=T_0$, thus are the same,
\begin{equation}
\label{eq:T0Def}
Z_{N,1, \rm kin}(\beta_0)e^{-\beta_0{E(N)}} = Z_{1,N, \rm kin}(\beta_0),\quad \beta_0\equiv\frac 1{k_{\rm B}T_0}.
\end{equation}
While the left-hand side is known exactly [$Z_{N,1, \rm kin}(\beta)$ is given by Eq.~(\ref{eq:Zonesoliton})], the right-hand side of Eq.~(\ref{eq:T0Def}) lies between the bounds given by Eq.~(\ref{eq:bounds}). Taking the $N$th root of Eq.~(\ref{eq:T0Def}) for each of these bounds leads~\cite{maple}, in the thermodynamic limit~(\ref{eq:limit}), to two characteristic, $N$-independent temperatures
\begin{align}
T_1^{(\infty)} &= \frac 23\frac{E_{\rm gap}}{k_{\rm B}} \frac {1}{W\left[\frac 83\pi B\exp(2)\right]},\\
T_2^{(\infty)} & = \frac 23\frac{E_{\rm gap}}{k_{\rm B}} \frac {1}{W\left[\frac 23\pi B\exp(2)\right]},
\label{eq:T2inftydef}
\end{align}
where $W(x)$ is the Lambert W function which solves $W(x)\exp[W(x)]=x$~\cite{maple}. In the thermodynamic limit~(\ref{eq:limit}), the temperature for which it is equally probable to find $N$ single particles and one bright soliton is lies in the range
\begin{equation}
\nonumber
0<T_1^{(\infty)} \le T_0^{(\infty)} \le T_2^{(\infty)} < \infty
\end{equation}
For  numerical finite-size investigations we focus on particle numbers $N\approx 100$ relevant for generation of Schr\"odinger-cat states on timescales shorter than characteristic decoherence times~\cite{WeissCastin2009};  $T_2^{(\infty)}$ turns out to be a  characteristic temperature scale already for these particle numbers (see Fig.~\ref{fig:transition}).

Figure~\ref{fig:transition} shows that the numerical data obtained via exact recurrence relations for the canonical partition function [Eqs.~(\ref{eq:ZNchosenconfig})-(\ref{eq:exactrec2})].
at the transition many solitons are involved (Fig.~\ref{fig:transition}~d). Near $T_2^{(\infty)}$, the numerical data is consistent with both the specific heat and the temperature-derivative of $\langle n_1\rangle$ scaling $\propto N^2$ for $N\approx 100$.

\begin{figure}
\includegraphics[width=\linewidth]{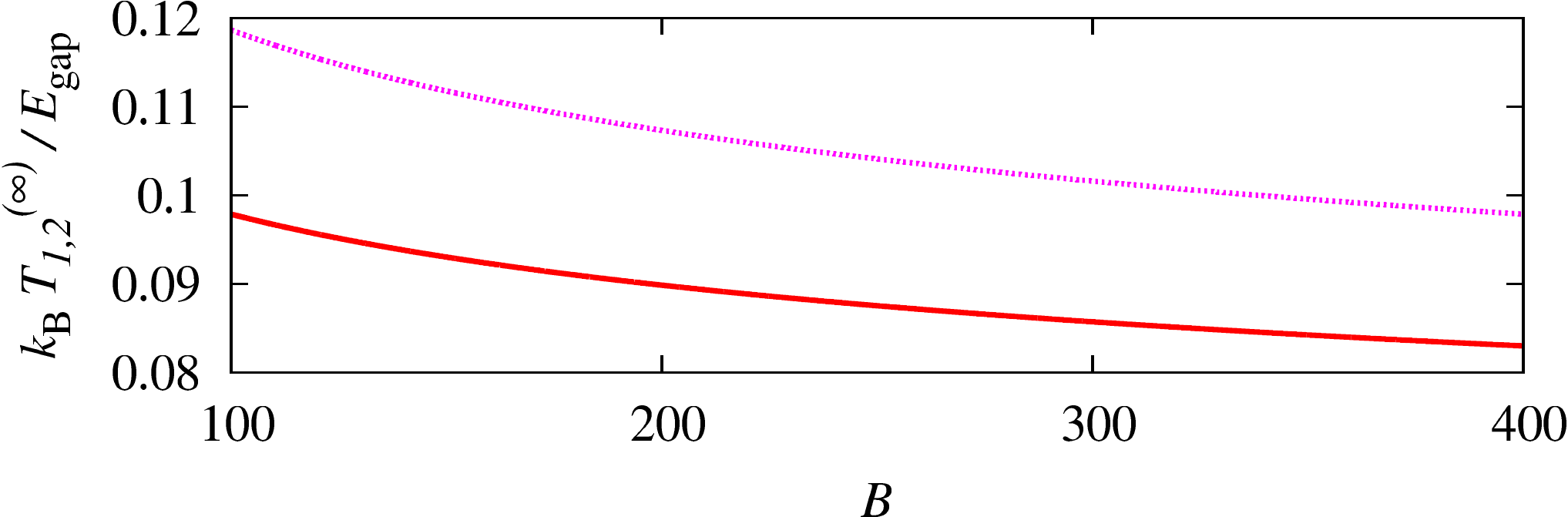}
\caption{\label{fig:transitionTemp} (Color online)
Characteristic temperatures for the soliton--non-soliton transition in the thermodynamic limit~(\ref{eq:limit}) as a function of $B$ [Eq.~(\ref{eq:ratio}) which depends on the ratio of the system size and the size of a two-particle soliton.]. Upper, magenta/gray dotted curve: above this temperature, the relative number of atoms in the largest soliton vanishes [Eq.~(\ref{eq:orderhighT})]. For temperatures below the red/black solid curve, a macroscopically occupied bright soliton exists [Eq.~(\ref{eq:orderlowT})]. The transition region is small compared to the difference of the two curves~(cf.~Fig.~\ref{fig:transition}).}
\end{figure}

To demonstrate that we indeed have a phase transition let us start by focusing on cases where we have $N-n$ particles in one soliton and $n$ free particles; $n=\mathcal{O}(N)$ and $N\ggg 1$. Using Eq.~(\ref{eq:T0Def}) to express the partition function for $n$ free particles corresponds to a system with fewer atoms ($n$) but the same $g_{\rm 1D}$ thus rescaling $E_{\rm gap}$ and therefore also  $T_{1,2}^{(\infty)}$  by a factor of $\frac {n^2}{N^2}$; the bounds in Eq.~(\ref{eq:bounds}) now become for not too low temperatures
\begin{align}
Z_{n,1,\rm kin}\left(\frac {1}{k_{\rm B}T_2^{(\infty)}}\right)\left(\frac{T}{T_2^{(\infty)}\frac {n^2}{N^2}}\right)^{\frac n2}\exp\left(\frac{(n+1)E_{\rm gap}}{3k_{\rm B}T_2^{(\infty)}}\right)
 \le Z
\nonumber\\\le
Z_{n,1,\rm kin}\left(\frac {1}{k_{\rm B}T_1^{(\infty)}\frac {n^2}{N^2}}\right)\left(\frac{T}{T_1^{(\infty)}\frac {n^2}{N^2}}\right)^{\frac n2}\exp\left(\frac{(n+1)E_{\rm gap}}{3k_{\rm B}T_1^{(\infty)}}\right),
\label{eq:bounds2}
\end{align}
with $Z=Z_{1,n, \rm kin}(\beta)$ and $T_2^{(\infty)}>T_1^{(\infty)}$. Multiplying this equation with $\exp[-\beta E_0(N-n)]$ to obtain the full partition function with $N-n$ particles in one soliton and $n$ free particles and dividing by $\exp[-\beta E_0(N)]$ yields that for $n\approx N$ the $n$-dependence (and in particular the question if they grow or shrink) is dominated by the $\left({T}/{T_{1,2}^{(\infty)}}\right)^{n/2}$-terms. Including factors of the order of~(\ref{eq:exgrowth}) to include the contribution of all other configurations with $N-n$ particles in one soliton (or directly including terms with more than one small soliton) does not change the convergence behavior.
Summing over $n\approx N$ such that the sum includes a finite fraction of $N$, say, all $n\ge0.99N$, we thus have
\begin{equation}
\label{eq:orderlowT}
0< \lim_{N\to \infty}\frac{\langle n_1\rangle}N \le 1, \quad T<0.99^2T_1^{(\infty)}.
\end{equation}
 Extending the above reasoning based on Eq.~(\ref{eq:bounds2}) to high temperatures ($T>T_2^{(\infty)}$) shows that in the sum~(\ref{eq:n1}): 
\begin{equation}
\label{eq:orderhighT}
 \lim_{N\to \infty}\frac{\langle n_1\rangle}N =0, \quad T> T_2^{(\infty)}.
\end{equation}
Thus, using the canonical ensemble~\cite{PathriaBeale2011} we have
shown the existence of a phase transition in the thermodynamic
limit~(\ref{eq:limit}) [cf.\ Fig.~\ref{fig:transitionTemp}].

 The $N$-dependence of the specific heat shows that both in the high-temperature phase [Fig.~\ref{fig:transition} (b),(c)] and in the low-temperature phase [Fig.~\ref{fig:transition} (b)] predictions of the canonical and the microcanonical ensemble~\cite{PathriaBeale2011} agree~\cite{PathriaBeale2011}. As the mean energy is a monotonously increasing function of temperature [Eq.~(\ref{eq:C})] and as furthermore, the choice of the thermodynamic limit~(\ref{eq:limit}) leads to a mean energy $\propto N$ and an $N$-independent temperature scale, the $\propto N^2$ behavior displayed by the specific heat in Fig.~\ref{fig:transition} (d) can only occur in a small ($\propto 1/N$) temperature range in which both ensembles no longer are equivalent.
 
To conclude, we find the existence of a finite-temperature
many-particle phase transition in a one-dimensional quantum
many-particle model, the homogeneous Lieb-Liniger gas with attractive
interactions [Eqs.~(\ref{eq:orderlowT}) and (\ref{eq:orderhighT});
Fig.~\ref{fig:transitionTemp}]. The low temperature phase consists of
a macroscopic number of atoms being one large quantum matter-wave
bright soliton  with delocalized center-of-mass wave function (which
does not display long-range order thus not violating the Mermin-Wagner
theorem~\cite{MerminWagner1966,Hohenberg1967};  the Landau
criterion~\cite{LandauLifshitz2002b} which argues against the
co-existence of two distinct phases is also not violated); the high temperature phase is a free gas. As a harmonic trap would facilitate soliton formation~\cite{BillamEtAl2012}, we conjecture that the existence of a finite-temperature phase transition remains true for weak harmonic traps. In experiments, even the \textit{integrable}\/ Lieb-Liniger gas can thermalize as the wave guides are \textit{quasi}-one-dimensional (cf.~\cite{MazetsSchmiedmayer2010,CockburnEtAl2011}).

Via exact canonical recurrence relations we also numerically investigate the experimentally relevant case of some 100 atoms (cf.~\cite{MedleyEtAl2014,WeissCastin2009,StreltsovEtAl2009b}) with the (experimentally realizable~\cite{SchmidutzEtAl2014}) box potential. The spike-like specific heat provides further insight: the specific heat ($\propto N^2$) is the derivative~(\ref{eq:C}) of an energy scaling not faster than $\propto N$~(\ref{eq:limit}).  At low temperatures all atoms form one soliton; the size of the soliton thus is an ideal experimental signature (cf.~\cite{KhaykovichEtAl2002,StreckerEtAl2002,CornishEtAl2006,MarchantEtAl2013,MedleyEtAl2014,McDonaldEtAl2014,NguyenEtAl2014,MarchantEtAl2016,EverittEtAl2015,LepoutreEtAl2016}).


\acknowledgments
I thank T.~P.~Billam, Y.~Castin, S.~A.~Gardiner, D.\ I.\ H.\ Holdaway, N.\ Proukakis and T.\ P.\ Wiles for discussions and the
UK EPSRC for funding (Grant No.~EP/L010844/1 and EP/G056781/1). 
The data presented in this paper
will be
available online~\cite{Weiss2016Data}.

\textit{Note added:}\/ Recently, a related work appeared~\cite{HerzogEtAl2014}.

%

\end{document}